\documentclass[reprint,
aps,superscriptaddress,twocolumn,showpacs,longbibliography]{revtex4-2}
\usepackage{graphicx,color}
\usepackage{amsmath,amsfonts,enumerate,amsthm,amssymb,bbm}
\usepackage[colorlinks=true,citecolor=blue,linkcolor=magenta]{hyperref}
\usepackage{multirow}
\usepackage{bbold}

\usepackage{braket}
\usepackage{soul}

\usepackage[caption = false]{subfig}

\usepackage{scrextend}
\usepackage{xcolor}
\usepackage{dsfont}
\usepackage{nicefrac}
\usepackage[linesnumbered,ruled,vlined]{algorithm2e}
\usepackage{quantikz}
\usepackage{apptools}

\mathchardef\ordinarycolon\mathcode`\:
\mathcode`\:=\string"8000
\begingroup \catcode`\:=\active
  \gdef:{\mathrel{\mathop\ordinarycolon}}
\endgroup

\theoremstyle{plain}

\theoremstyle{definition}

\theoremstyle{remark}

\usepackage{amsmath,bm}
\usepackage{todonotes}

\AtAppendix{\counterwithin{thm}{section}}
\AtAppendix{\counterwithin{corol}{section}}
\AtAppendix{\counterwithin{lem}{section}}
\AtAppendix{\counterwithin{propos}{section}}
\AtAppendix{\counterwithin{defn}{section}}
\AtAppendix{\counterwithin{rmk}{section}}
\AtAppendix{\counterwithin{ex}{section}}

\def\>{\rangle}
\def\<{\langle}

\renewcommand{\ket}[1]{|#1\rangle}               
            
\renewcommand{\bra}[1]{\langle #1|}

\renewcommand{\vec}[1]{\boldsymbol{#1}}

\begin{document}
\date{\today}
\title{Quantum Carleman linearisation efficiency in nonlinear fluid dynamics}

\author{Javier Gonzalez-Conde}
\email[Corresponding author: ]{\qquad javier.gonzalezc@ehu.eus}

\affiliation{Department of Physical Chemistry, University of the Basque Country UPV/EHU, Apartado 644, 48080 Bilbao, Spain}
\affiliation{EHU Quantum Center, University of the Basque Country UPV/EHU, Apartado 644, 48080 Bilbao, Spain}
\affiliation{Quantum Mads, Uribitarte Kalea 6, 48001 Bilbao, Spain}

\author{Dylan Lewis}
\affiliation{Department of Physics and Astronomy, University College London, London WC1E 6BT, United Kingdom}
\author{Sachin S. Bharadwaj}
\affiliation{Department of Mechanical and Aerospace Engineering, New York University, New York 11201 USA}

\author{Mikel Sanz}

\affiliation{Department of Physical Chemistry, University of the Basque Country UPV/EHU, Apartado 644, 48080 Bilbao, Spain}
\affiliation{EHU Quantum Center, University of the Basque Country UPV/EHU, Apartado 644, 48080 Bilbao, Spain}
\affiliation{IKERBASQUE, Basque Foundation for Science, Plaza Euskadi 5, 48009, Bilbao, Spain}
\affiliation{Basque Center for Applied Mathematics (BCAM), Alameda de Mazarredo, 14, 48009 Bilbao, Spain}

\begin{abstract}
Computational fluid dynamics (CFD) is a specialised branch of fluid mechanics that utilises numerical methods and algorithms to solve and analyze fluid-flow  problems. One promising avenue to enhance CFD is the use of quantum computing, which has the potential to resolve nonlinear differential equations more efficiently than classical computers. Here, we try to answer the question of which regimes of nonlinear partial differential equations (PDEs) for fluid dynamics can have an efficient quantum algorithm. We propose a connection between the numerical parameter, $R$, that guarantees efficiency in the truncation of the Carleman linearisation, and the physical parameters that describe the fluid flow. This link can be made thanks to the Kolmogorov scale, which determines the minimum size of the grid needed to properly resolve the energy cascade induced by the nonlinear term. Additionally, we introduce the formalism for vector field simulation in different spatial dimensions, providing the discretisation of the operators and the boundary conditions.

\end{abstract}

\maketitle
\section{Introduction}
    
Fluid-flow equations play a crucial role in various scientific and engineering disciplines, serving as a foundational framework to  analyze fluid-flow  behavior.  The properties and solutions of fluid-flow equations are indispensable for optimizing the design of systems such as aircrafts, pipelines, and hydraulic structures, ensuring their efficiency, stability, and safety~\cite{aerodynamic, Hydraulic, pipeline}. These equations serve as a mathematical framework articulating essential parameters—namely, velocity, pressure, and density distributions—comprehensively characterizing the dynamic behavior of fluids~\cite{Zamir2000, Serrin1959, MARKATOS1986190, LAUNDER1974269, Solomon2009}. Beyond the primary variables, these equations might also include the modeling of lift and drag, which incorporates a bias or preferred direction in the evolution of the velocity field, and turbulence~\cite{MARKATOS1986190, LAUNDER1974269, Solomon2009,iyer2021area}, which can arise in nonlinear partial differential equations (PDEs) and is characterized by chaotic evolution in both space and time.

The inherent nonlinearity in these equations poses challenges for obtaining analytical solutions, particularly when dealing with turbulence and complex boundary conditions.
In this sense, computational fluid dynamics (CFD) \cite{anderson1995computational, chung2002computational} is a specialized branch of fluid mechanics that utilizes numerical methods and algorithms to solve and analyze fluid-flow  problems, leveraging computational power to simulate and visualize the behavior of fluids in complex systems. However, when solving such nonlinear differential equations, these techniques are limited by the scaling of the required computational cost ~\cite{10.3389/fphy.2013.00002, pires2022challenges, Spalart_Venkatakrishnan_2016}, which rapidly exceeds the capacity of traditional computers to tackle nonlinearity.

A promising avenue to enhance CFD is the use of quantum computing \cite{Yepez2002,PhysRevE.63.046702,Yepez1999,
yepez2002efficientquantumalgorithmonedimensional,Mezzacapo2015,steijl2018parallel,Somma2017,ray2019solvingnavierstokesequationquantum,Navier_PDE,Budinski2021,Moawad2022,app12062873,PhysRevResearch.4.033176,itani2023quantumalgorithmlatticeboltzmann,
Sachin_NATO,syamlal2024computationalfluiddynamicsquantum,succi2023quantumcomputingfluidsstand,bharadwaj2024simulating, chen2021quantumfinitevolumemethod,  wu2024quantumalgorithmsnonlineardynamics,bharadwaj2020quantum, Kocherla_2024} which has the potential to handle complex problems more efficiently than classical computers for certain types of calculations \cite{yepez1998quantum,jaksch2023variational, Lubasch_2020, chen2022quantum, lewis2023limitations, ingelmann2024two, bharadwaj2023hybrid,lu2023quantum, meng2024simulating, Rene_QAfF,Rene_QAfF_2,  Sanavio_2024, tennie2024quantumcomputingnonlineardifferential, su2024quantumstatepreparationvelocity, sanavio2024carlemangradapproachquantumsimulation, penuel2024feasibilityacceleratingincompressiblecomputational, costa2023improvingquantumalgorithmsnonlinear, succi2024carlemanroutesquantumsimulation,bharadwaj2024qflows}. Recently, a variety of quantum algorithms designed for addressing  ordinary  differential equations (ODEs) \cite{Berry_ODEs, Berry_ODEs_2, Xin_ODEs, Chils_ODEs, Costa_ODEs, Dong_ODEs,Fang2023timemarchingbased, berry2022quantum, jennings2024cost} and PDEs \cite{Childs_PDEs, Arrazola_PDEs, Montanaro_PDEs, Alan_PDEs, Xue_2021, Sachin_homotopy, Liu_2021, an2022efficient, Krovi2023improvedquantum,bharadwaj2024compact} has emerged within the scientific literature, proposing algorithms to solve the Poisson equation \cite{Poisson_PDE,bharadwaj2023hybrid}, the Dirac equation \cite{Dirac_PDE}, the heat equation \cite{Heat_PDE}, the wave equation \cite{Suau_2021}, the Black-Scholes equation \cite{ PhysRevResearch.5.043220}, the Vlasov equation \cite{Vlasov_PDE} or the Navier-Stokes equation \cite{Navier_PDE}, among others. 

When solving nonlinear PDEs, numerical schemes typically introduce a linearisation of the equation \cite{Xue_2021, Sachin_homotopy, Liu_2021, an2022efficient,  Krovi2023improvedquantum,lewis2023limitations}, which produces a controllable error. The resulting linear system of equations is subsequently solved with a quantum linear systems algorithm~\cite{HHL,ambainis_HHL, Childs_HHL, Wossnig_HHL, Somma_HHL,Clader_HHL,Huang_HHL, Costa_HHL, jennings2023efficient}, which potentially introduces an exponential advantage over classical solvers. One such methodology to linearise PDEs employs the homotopy perturbation method, as outlined in the work by Xue et al.~\cite{Xue_2021, Sachin_homotopy} to solve homogeneous nonlinear dissipative ODEs. Alternatively, some recent proposals in the literature suggest the use of the quantum Carleman linearisation (QCL) algorithm~\cite{Liu_2021, an2022efficient, Krovi2023improvedquantum}, which maps a finite dimensional nonlinear system to an infinite-dimensional linear system. By truncating the linear system with a controllable error, quantum algorithms for solving linear systems of equations can potentially offer an exponential speedup in obtaining numerical solutions to PDEs.

The sufficient condition to reach an exponential error suppression in the truncation relies on a certain parameter, $R$ (defined in Section~\ref{sec:numerical_methods}), being smaller than 1. However, as the number of grid points increases for a fixed PDE, the parameter $R$ also increases. Indeed, $R$ can grow to the point that the condition  $R<1$~\cite{Liu_2021} is no longer fulfilled, despite the fact the physics of the system has not changed.

Here, we propose a connection between the numerical parameter, $R$, that guarantees the efficiency of the QCL algorithm for solving a fluid PDE with the physical parameters that describe the fluid flow  in different spatial dimensions. The resolution limit to resolve all the physics of the system is determined by physical length scales \cite{iyer2021area}, which provide a sufficiently fine  discretisation to capture the entire energy cascade induced by the nonlinear term. Thus, since the effect of the nonlinear term on the solution is already resolved, all the physics of the system can be captured. In this sense, we are assuming that the hypothetical \textit{``continuous solution''} does not provide additional information of the dynamics beyond the smallest physical length scale. Furthermore, our contribution extends the formalism for vector field simulation of fluid-flow  equations in an arbitrary dimension.

The article is structured as follows: First, in Section \ref{sec:fluids}, we introduce fluid-flow  equations. Then, in Section \ref{sec:discret} we present the problem setting and subsequently, in Section \ref{sec:numerical_methods}, the numerical methods to solve this set of PDEs on quantum computers, carefully detailing the resources needed according to the length scales present in the different dimensions. Finally, in Section \ref{sec:num_results}, we provide some numerical results to illustrate our work.

\section{\label{sec:fluids}Fluid flows}
For a continuum-level model of general fluid-flow  problems, the Navier-Stokes (NS) equations  capture the actual physics of general fluid flows \cite{constantin1988navier}. Nevertheless, one could also consider alternative formulations \cite{sanavio2024carlemangradapproachquantumsimulation}. 
In this work we confine our consideration to scenarios in which the fluid dimension is restricted to the values $d=1,2,3$. Furthermore, our emphasis will be on incompressible flows with uniform viscosity. The general form of the NS equations that describe the  velocity field $\bm{u}(\bm{x}, t) =~ (u_1(\bm{x},t),  ..., u_d(\bm{x},t))$ of an incompressible fluid flow  reads
 \begin{align}
    \frac{\partial \textbf{u}}{\partial t} + \textbf{u}\cdot\nabla \textbf{u} &= \nu \nabla^{2}\textbf{u} + \bm{f}(\bm{x}, t),
    \label{eq:NS1}\\
    \nabla\cdot\textbf{u} &= 0.
    \label{eq:NS2}
\end{align}
where  $\bm{x} = (x_1, x_2, \dots, x_d)\in \mathbb{R}^d$ is the $d$-dimensional coordinate vector, $\bm{f}(\bm{x}, t)=~ - \frac{1}{\rho}\nabla p + \mathbf{g/\rho}$, with $p$ the pressure, $\mathbf{g}$ the forcing/stirring term, and $\nu$ is the kinematic viscosity.

\begin{figure}[t!]
    \centering
    \includegraphics[width=.85\columnwidth]{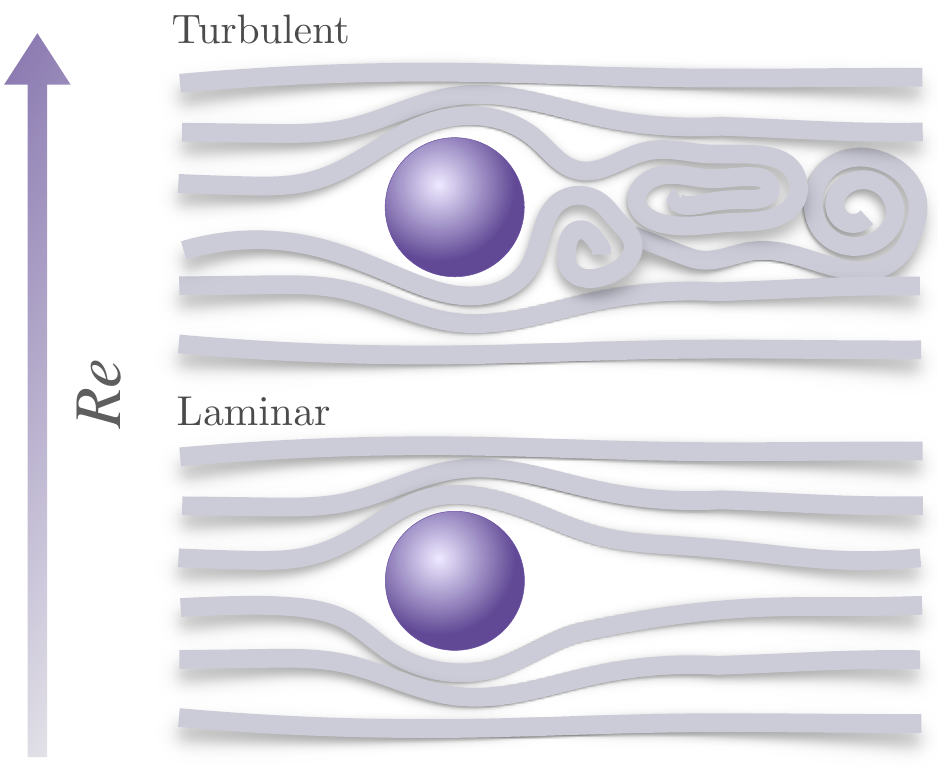}
    \caption{ At low Reynolds numbers, fluid flow tends to be dominated by a laminar regime, where the motion of the fluid is smooth and orderly, with layers of fluid sliding past each other in a stable fashion. In this regime, viscous forces play a significant role in maintaining the flow's structure. However, as the Reynolds number increases, the flow transitions into a turbulent regime, characterized by chaotic, irregular motion and the formation of eddies and vortices. In turbulent flows, inertial forces become dominant, leading to complex fluid behavior.}
    \label{fig:turbulence_regime}
\end{figure}

A quantity of particular interest when studying fluids is the the Reynolds number \cite{wei1989reynolds}, $Re$, which we define and study in detail in Section \ref{sec:len_scale}. This dimensionless quantity helps to predict fluid-flow  patterns in various situations by measuring the ratio between inertial and viscous forces. At low Reynolds numbers, flows tend to be laminar, while at higher Reynolds numbers they transition into a turbulent flow, see Fig.~\ref{fig:turbulence_regime}.

\section{Problem Setting}
\label{sec:discret}

Let us now present the problem setup to be numerically solved on a quantum computer.

\subsection{Spatial discretisation}

Numerically solving fluid-flow  equations numerically requires a discretisation of both space and time. Here, in order to provide explicit bounds for the operators, we present a formalism to discretise Eq.~\eqref{eq:NS1}. This formalism can be easily extrapolated to the general case. We assume that the solution domain is a hypercube, $\Omega \in \mathbb{R}^d$, of side length $\mathcal{L}$ with dimension $d$. We discretise space such that every coordinate of $\bm{x}\in \mathbb{R}^d$ is only defined at $N$ discrete points, where we assume $N$ to be a power of 2. Note that in this work, we chose a homogeneous spatial discretisation, but this might not apply in general.  Following this, both the driving term $\bm{f}(\bm{x},t)$ and the field solution $\bm{u}(\bm{x}, t)$ become  $dN^d$-dimensional vectors,  $\bm{F_0}(t)$ and $\bm{U}(t)$, respectively. This last vector can be written as 
\begin{small}
\begin{equation}
  \bm{U}(t)=\big(\underbrace{ U_1^{(1)}\dots\ U_{1}^{(N^d)}}_{U_1} U_2^{(1)}\dots\ U_{2}^{(N^d)} \dots \ U_d^{(1)}\dots\ U_{d}^{(N^d)}\big)^T.
\end{equation}
\end{small}
As in Eq. \eqref{eq:NS1}, the spatially-discretised  nonlinear system of a typical fluid-flow  equation is quadratic in the nonlinearity, and it is of the form

\begin{align}
    \frac{d \bm{U}(t)}{d t}= F_2 \bm{U}(t)^{\otimes 2}+ F_1 \bm{U}(t) + F_0(t),
    \label{eq_setup}
\end{align}
where we define
\begin{equation}
    F_1 =  \nu \sum_{l=1}^d \Bigg( \bigoplus_{i=1}^{l-1}O \oplus     \overbrace{\sum_{ \bm{j} \in \Delta}\bm{D}^{\bm{j}}}^\mathfrak{D}  \bigoplus_{i=l+1}^{d}O\Bigg),
    \label{eq:f1}
\end{equation}
with $\Delta=\{ (2,0, \dots ,0), (0,2, \dots ,0)\dots (0,0, \dots ,2)\}$, the multidimensional derivative operator $\bm{D}^{\bm{j}}=D^{j_1}\otimes D^{j_2} \otimes \dots \otimes D^{j_d}$, $\bm{j}~=~(j_1, j_2, \dots, j_d)$ with all $j_i \in \mathbb{Z}^{0+}$, $O$ the $N^d \times N^d $ null matrix and $D$ the single coordinate derivative operator (which can be defined in a number of ways as discussed in Section~\ref{sec:boundary_conditions}) and
\begin{multline}
    F_2 = - \sum^d_{l=1}   \sum^d_{q=1} \bm{P} \Bigg(\left\{ \bigg(\bigoplus_{i=1}^{l-1}O\oplus\mathds{1}  \bigoplus_{i=l+1}^{d}O\bigg) \right\} \otimes \\ \left\{ C^{q-l} \bigg( \bigoplus_{i=1}^{l-1}O \oplus \bm{D}^{\hat{{e}}_q} \bigoplus_{i=l+1}^{d}O \bigg) C^{l-q} \right\}\Bigg),   
     \label{eq:f2}
\end{multline}
 where  $\hat{e}_q$ is the unit vector in the  $q$-th coordinate and $\bm{P}\in~ \mathbb{R}^{d^2N^{2d}\times d N^d}$ is the projector matrix that takes a vector from $d^2N^{2d}$ dimensions to $dN^{d}$,
\begin{align}
    \bm{P} = \begin{pmatrix}
        \delta_{11} & \delta_{22} & \dots & \delta_{dN^d dN^d}
    \end{pmatrix},
\end{align}
with the $dN^d \times dN^d$ projector $\delta_{ii}=\ket{i}\bra{i} $,
and the cyclic permutation matrix 
\begin{align*}
    C = \begin{pmatrix}
        \bm{0} & \mathds{1}_{N^d} & \bm{0} & \cdots & \bm{0} \\
        \bm{0} & \bm{0} & \mathds{1}_{N^d} & \ddots & \vdots \\ \
        \vdots &  & \ddots & \mathds{1}_{N^d} &  \\
        \mathds{1}_{N^d} & \cdots & \ddots & \ddots &  \\
    \end{pmatrix},
\end{align*}
which is composed of blocks of identities and satisfies $C^{d}=~\mathds{1}_{dN^d}$, meaning $C^{-g}$ with $g\in\mathbb{Z}^{0+}$ is equal to $ C^{d-g}$. Note that both $F_2 \in \mathbb{R}^{dN^d\times d^2N^{2d}}$ and $F_1 \in \mathbb{R}^{dN^d\times dN^d}$ are time-independent while $F_0(t) \in \mathbb{R}^{dN^d}$ is the time-dependent driving term. Even though the specific form of $F_1$ and $F_2$ follows from the choice of the derivative operator $D$, which depends on the  discretisation and boundary conditions, we can depict some properties of the two operators in general. 

First, we note that, by construction, if the derivative operator defined on a single variable, $D$, is diagonalisable, then $F_1$ is diagonalisable. Additionally, the choice of $D$ for every spatial coordinate, which is strictly tied to the boundary conditions, will determine the spectrum of the matrix $F_1$. Hence, the spectrum of the operator $\mathfrak{D}=~\sum_{\bm{j} \in \Delta} \bm{D}^{\bm{j}}$ is given by the sum of the spectra of every $\bm{D}^{\bm{j}}$. If we can guarantee that the real part of the eigenvalues of the spectrum of the operator $\mathfrak{D}$ is negative, then, as $F_1$ is composed of direct sums of $\mathfrak{D}$ (see Eq. \ref{eq:f1}), we straightforwardly obtain that $F_1$ is strictly negative. On the other hand, we can upper bound the magnitude of the nonlinear operator $F_2$ in terms of the spectral norm
\begin{align}
    \Vert F_2 \Vert_2 &\leq \Vert \bm{P} \Vert_2 \Bigg\Vert   \sum^d_{l=1}   \sum^d_{q=1}  \Bigg(\left\{ \big(\bigoplus_{i=1}^{l-1}O \oplus\mathds{1}  \bigoplus_{i=l+1}^{m}O\big) \right\} \nonumber \\ 
    & \otimes  \left\{ C^{q-l} \big( \bigoplus_{i=1}^{l-1}O \oplus \bm{D}^{\hat{{e}}_q} \bigoplus_{i=l+1}^{m}O \big) C^{l-q} \right\}\Bigg) \Bigg\Vert_2 \nonumber \\  
    &\leq \sum^d_{l=1}   \sum^d_{q=1}   \Big\Vert  \bm{D}^{\hat{{e}}_q}   \Big\Vert_2 =   \sum^d_{l=1}   \sum^d_{q=1}   \big\Vert  D  \big\Vert_2  = d^2  \big\Vert  D  \big\Vert_2, 
\end{align}
where we have used that the magnitude of the projector $\Vert \bm{P} \Vert_2=1$.  

\subsection{Boundary conditions\label{sec:boundary_conditions}}
In order to define the derivative operators on a single coordinate, $D$, for different boundary conditions, we assume a homogeneous discretisation. The grid point spacing is therefore $\Delta x = \frac{\mathcal{L}}{N-1}$. The derivatives are computed with the second order central finite difference method \cite{LI200529}, which leads to a sparse representation of $F_2$ and $F_1$. 
Thus, the discretised definition of these two operators, including the choice of boundary condition, transforms the nonlinear system of PDEs into a system of nonlinear ordinary differential equations (ODE). We now study various possible choices of boundary conditions and how they determine the single coordinate derivative operator.

\subsubsection{Periodic boundary conditions}

The single coordinate derivative operator with periodic boundary conditions (PBC) can be written as
\begin{align}
\label{eq:BPBC}
    D_\text{PBC} = \frac{1}{2 \Delta x} \begin{pmatrix}
        0 & 1 & 0 &  \dots & -1 \\
        -1 & 0 & 1 &  \ddots  & \vdots\\
        0 & -1 & 0 & \ddots &  0 \\
        \vdots & \ddots & \ddots & \ddots & 1  \\
        1 & \dots & 0 & -1 & 0  \\
    \end{pmatrix},
\end{align}
with eigenvalues 
$$\lambda^{(k)}_{D_\text{PBC}}=i\sin\left(\frac{2\pi k}{N}\right)/(2\Delta x), \ \ \ k=0 \ ... \ N-1.$$ 
Higher order derivatives can be defined as powers of the first order derivative operator. Note that all the eigenvalues are purely imaginary but $\lambda^{(0)}=\lambda^{(N/2)}=0$, which corresponds to the subspace spanned by the vectors $\frac{1}{\sqrt{N}}(1\ 1\ 1\ \dots 1\ 1\ 1)^T$ and $\frac{1}{\sqrt{N}}(1\ -1\ 1\ \dots 1\ -1\ 1)^T$. Additionally, the spectral norm of $D_\text{PBC}$ can be calculated as $||D_\text{PBC}||_2=\frac{N-1}{2\mathcal{L}}$. 

In this set up, the eigenvalues of the operator $\bm{D}^{{{2\hat{e}_l}}}_\text{PBC}$ for $\bm{j} = 2\hat{e}_l$, where $\hat{e}_l$ is the unit vector in the dimension $l$, 
are
\begin{small}
$$\lambda_{\bm{D}^{2 \hat{e}_l}}^{(k)}=-\frac{1}{(2\Delta x)^2}\sin^2\left(\frac{2\pi ((k \text{ mod } N^{d}/N^l)\text{ mod } N)}{N}\right),  $$
\end{small}
with $k=1\ \dots\ N^d$. Therefore, as $\{\bm{D}^{\bm{j}}_\text{PBC}\}_{ \bm{j} \in \Delta}$ is a set of commuting observables,  the eigenvalues of  $\sum_{ \bm{j} \in \Delta}  \bm{D}^{\bm{j}}_\text{PBC}$ are the sum of the eigenvalues of every $\bm{D}^{\bm{j}}_\text{PBC}$
\begin{small}
$$\footnotesize{\lambda_{\sum_{ \bm{j} \in \Delta}  \bm{D}^{\bm{j}}}^{(k)}=\frac{-1}{(2\Delta x)^2}\sum_{l=1}^d \sin^2\left(\frac{2\pi ((k \text{ mod } N^{d}/N^l)\text{ mod } N)}{N}\right).}  $$ 
\end{small}
Consequently, if all the coordinates have periodic boundary conditions, then $F_1$ has at least one zero eigenvalue, which means that the requirements for efficiency in Carleman linearisation are no longer fulfilled \cite{Liu_2021}. This issue might be circumvented if the the initial condition has no components in the subspace of the zero eigenvalue. Alternatively, one could also choose a non-homogeneous grid to tackle this issue. The first non-zero eigenvalue can be lower bounded by $$\frac{\sin^2\left(\frac{2\pi}{N}\right)}{(2\Delta x)^2}\sim \mathcal{O}\left(\frac{\pi^2}{\mathcal{L}^2}\right).$$

\subsubsection{Open boundary conditions}
The derivative operator with open boundary conditions (OBC) can be written as 
\begin{align}
\label{eq:BOBC}
    D_{\text{OBC}} = \frac{1}{2 \Delta x} \begin{pmatrix}
        0 & 1 & 0 &  \dots & 0 \\
        -1 & 0 & 1 &  \ddots  & \vdots\\
        0 & -1 & 0 & \ddots &  0 \\
        \vdots & \ddots & \ddots & \ddots & 1  \\
        0 & \dots & 0 & -1 & 0  \\
    \end{pmatrix}.
\end{align} 
Higher order derivatives can be defined as powers of the first order derivative operator.
As this is a tridiagonal symmetric matrix \cite{KULKARNI199963}, its  purely imaginary eigenvalues are

\begin{equation}
    \lambda^{(k)}_{D_{\text{OBC}}}= i \frac{(N-1)}{2\mathcal{L}}\cos\Big(\frac{\pi k }{N+1}\Big).
\end{equation}
The largest eigenvalue of $D_{\text{OBC}}^2$ is therefore 
\begin{equation}
    \lambda^{(N/2)}_{D_{\text{OBC}}^2}=-\left(\frac{(N-1)}{2\mathcal{L}}\right)^2\cos\Big(\frac{N \pi}{2(N+1)}\Big)^2.
\end{equation}
Thus, the largest eigenvalue of $F_1$ when all the coordinates have open boundary conditions is $$\nu d\lambda^{(N/2)}_{D_{\text{OBC}}^2}\sim \mathcal{O}(\nu 
 d\pi^2/\mathcal{L}^2).$$
Additionally, in order to calculate the norm of $D_{\text{OBC}}$, the eigenvalue  with the largest absolute value is 
\begin{equation}
    \lambda^{(1)}_{D_\text{OBC}}=i\frac{(N-1)}{2\mathcal{L}}\cos\Big(\frac{\pi}{N+1}\Big),
\end{equation} and thus
\begin{align}
    \Vert D_{\text{OBC}} \Vert_2 \leq \frac{(N-1)}{2\mathcal{L}}\cos\Big(\frac{ \pi}{(N+1)}\Big)\sim \mathcal{O}(N/\mathcal{L})
\end{align} 

\subsubsection{Dirichlet boundary conditions}
In this case, we assume that there exist fixed boundary values for $U_0(t), U_{N+1}(t)$, so that only the internal points are included in the discretisation. The derivative is therefore

\begin{align}
\footnotesize
     D_{\text{Dir}} U  = \frac{1}{2 \Delta x}
     \overbrace{\begin{pmatrix}
        0 & 1 & 0 &  \dots & 0 \\
        -1 & 0 & 1 &  \ddots  & \vdots\\
        0 & -1 & 0 & \ddots &  0 \\
        \vdots & \ddots & \ddots & \ddots & 1  \\
        0 & \dots & 0 & -1 & 0  \\
        \end{pmatrix}}^{\mathcal{D}} \begin{pmatrix}
        U_1   \\
        \vdots   \\
        \vdots   \\
        \vdots    \\
        U_N  \\
        \end{pmatrix}+\overbrace{\begin{pmatrix}
        -\frac{U_0(t)}{2\Delta x}  \\
        0  \\
        \vdots   \\
        0    \\
        \frac{U_{N+1}(t)}{2\Delta x}   \\
        \end{pmatrix}}^{\vec{b}}.
\end{align} 
Note that this formalism introduces an extra term given as a vector $\vec{b}$ which modifies the $F_0$ term, while the matrix $\mathcal{D}$ is the same as that in Eq. (\ref{eq:BOBC}). This prevents us from defining the second derivative operator as the square of the first order derivative. However, it can be defined as
\begin{align}
\footnotesize
     D^2_{\text{Dir}} U  = \frac{1}{\Delta^2 x}
     \overbrace{\begin{pmatrix}
        -2 & 1 & 0 &  \dots & 0 \\
        1 & -2 & 1 &  \ddots  & \vdots\\
        0 & 1 & 2 & \ddots &  0 \\
        \vdots & \ddots & \ddots & \ddots & 1  \\
        0 & \dots & 0 & -1 & 2  \\
        \end{pmatrix}}^{\mathcal{D}_2}\begin{pmatrix}
        U_1   \\
        \vdots   \\
        \vdots   \\
        \vdots    \\
        U_N  \\
        \end{pmatrix} +\overbrace{\begin{pmatrix}
        \frac{U_0(t)}{\Delta^2 x}   \\
        0  \\
        \vdots   \\
        0    \\
        \frac{U_{N+1}}{\Delta^2 x}(t)  \\
        \end{pmatrix}}^{\vec{b}_2}.
\end{align} 

Let us now detail some fundamental aspects of the operators $F_1$ and $F_2$ when Dirichlet conditions are assumed. When in comes to be an $N^d$-dimensional vector, the $D^2_{\text{Dir}}$ operator acting on the mathematical subsystem $l$ can be written as $$\bm{D}^{2 \hat{e}_l}U=(\mathds{I} \otimes \dots \otimes  \underbrace{\mathcal{D}_2}_{j_l}\otimes  \dots \otimes \mathds{I})U + (\textbf{1} \otimes \dots \otimes  \vec{b}_2 \otimes  \dots \otimes \textbf{1})$$ where $\textbf{1}$ is a vector of ones of length $N$. Therefore, the first part contributes to $F_1$ while the second term contributes to $F_0$. For the $F_2$ operator in Eq.~\eqref{eq:f2} we have a similar case, a first contribution to $F_2$ and a second additional contribution to $F_1$.    

\section{Numerical Methods\label{sec:numerical_methods}}

 Here we present the numerical scheme to solve the fluid-flow  equations on a quantum computer, where by ``solving'' a PDE using a quantum computer we mean obtaining an approximation to a quantum state whose amplitudes are proportional to the discretised solution vector.
\subsection{Carleman Linearisation}

The QCL algorithm uses a linearisation technique---Carleman embedding~\cite{Carleman}--- to transform a system of nonlinear differential equations  into a truncated linear system. For certain systems and initial conditions, the solutions behave such that the error from truncation can be suppresed exponentially fast by linearly increasing the truncation order, $C$ . The solution of the truncated linear differential equation can be found by using the quantum linear systems algorithm (QLSA)~\cite{HHL,ambainis_HHL, Childs_HHL, Wossnig_HHL, Somma_HHL,Clader_HHL,Huang_HHL, Costa_HHL} with exponential advantage.

Assuming that $F_1$ is diagonalisable and the real part of its eigenvalues are $\textrm{Re}(\lambda_N) \leq \dots  \leq \textrm{Re}(\lambda_1) < 0$, the sufficient condition for the efficiency of Carleman linearisation for any simulation time is given by the effective ratio of nonlinearity to linearity \cite{Liu_2021, Krovi2023improvedquantum} defined as
\begin{align}
    \label{eq:R_definition}
    R = \frac{1}{\vert\textrm{Re}(\lambda_1)\vert}\left(  \Vert u(0)\Vert_2 \Vert F_2\Vert_2  + \frac{\Vert F_0 \Vert_2}{\Vert u(0) \Vert_2}\right)<1,
\end{align} where $\bm{u}_0 = \bm{u}(0)$ is the initial condition at $t=0$, and $\Vert F_0(t)\Vert_2$ is the largest value of the spectral norm in the time interval $t=[0,T]$.  In Ref.~\cite{Krovi2023improvedquantum}, the real part of the largest eigenvalue of $F_1$ was replaced  by the log-norm of $F_1$.
As the proof given in Ref.~\cite{Liu_2021} does not depend on the spatial dimension $d$, if we can guarantee that $F_1$ is strictly negative for $d > 1$, we can extend for multidimensional operators the same sufficient condition on  $R$ to efficiently truncate the Carleman linearisation.

\begin{figure}[t!]
    \centering
    \includegraphics[width=1\columnwidth]{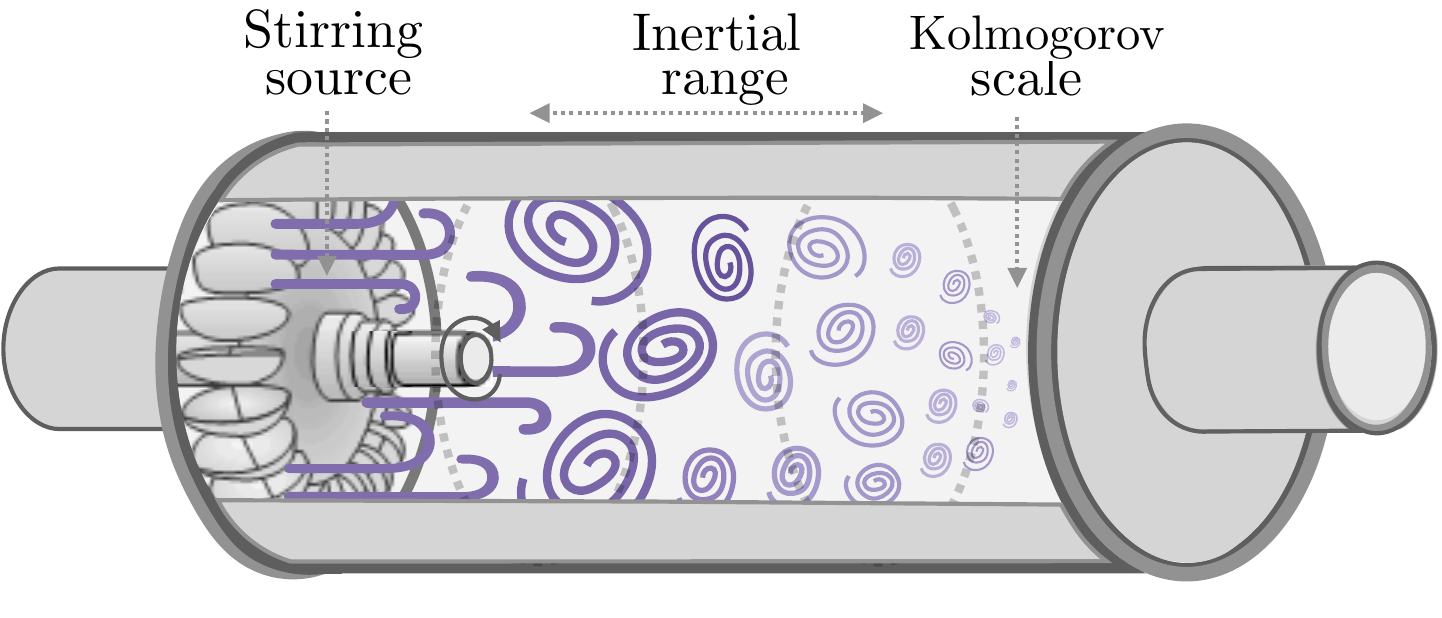}
    \caption{In a standard DNS setup, the flow is continuously energized at a large length scale. When the Reynolds number \(Re\) is sufficiently high, the flow becomes unstable and transitions into turbulence. This turbulence persists as long as the forcing continues; otherwise, it dissipates. The flow develops a wide range of scales, with much smaller structures than the stirring scale. Between these large and small scales lies the \textit{inertial range}, which is relatively independent of the extremes. Energy from large scales redistributes across smaller scales, where viscous forces dominate around structures with \(Re \sim 1\), leading to energy dissipation. The scale where this occurs is the Kolmogorov scale, \(\eta\), which is much smaller than the scale of the forcing term at high overall \(Re\).
}
    \label{fig:turbine}
\end{figure}

\begin{figure*}[t!]
    \centering
    \includegraphics[width=1.5\columnwidth]{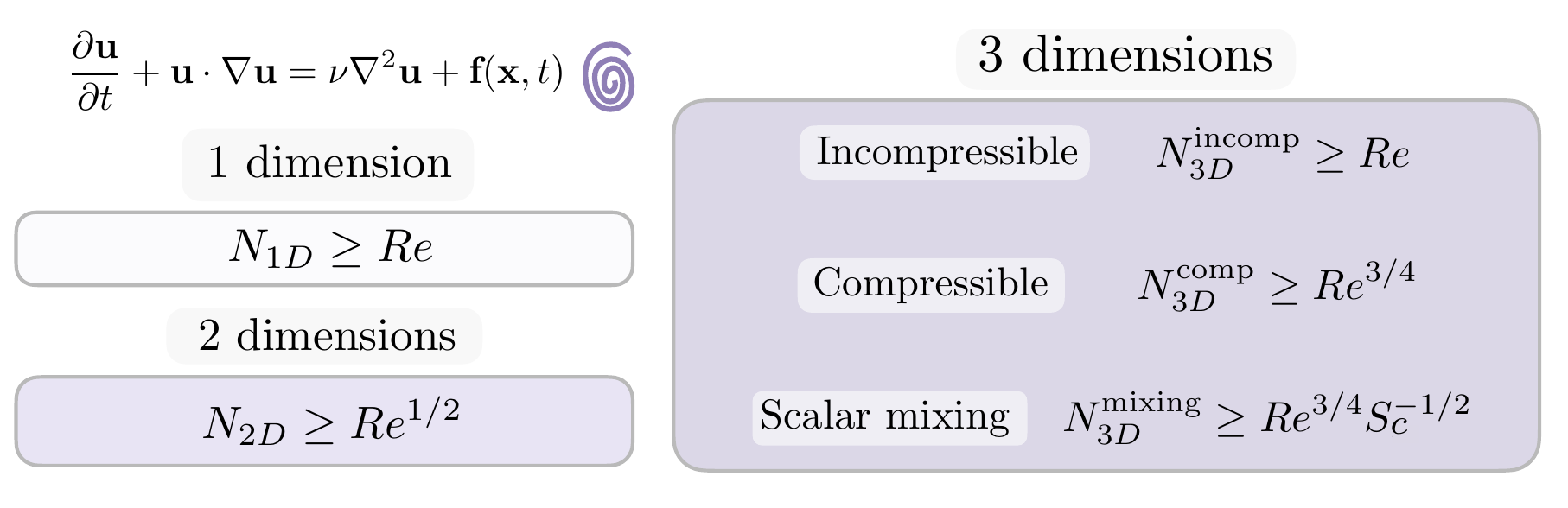}
    \caption{Summary of the minimum number of grid points needed per axis in order to resolve Kolmogorov length scales in 1, 2 and 3 dimensions assuming we can neglect boundary effects. In the case of 3 dimensions we can distinguish the incompressible regime, the compressible regime and the Turbulent passive scalar mixing regime. $Re$ denotes the Reynolds number and $Sc = \nu/D$ the Schmidt number, where $D$ is the molecular diffusivity coefficient and $\nu$ the kinematic viscosity.}
    \label{fig:Summary_dimensions}
\end{figure*}

Across all dimensions, a shared issue emerges: the definitions of \( R \), the eigenvalues of \( F_1 \), and the norms of \( F_0 \) and \( F_2 \), have a significant shortcoming. These values depend on the choice of the discretisation used to solve the continuous PDE. Thus, whether or not a provably efficient quantum algorithm exists may be conditioned on the discretisation size rather than the physical properties of the solutions. For instance, with open boundary conditions, following Section \ref{sec:boundary_conditions}, we have
\begin{align}
    \label{eq:R_definition}
    R(N) \approx \frac{\mathcal{L}^2}{d\nu\pi^2}\left(  \Vert u(0) \Vert_{L^2} \frac{d^2 N^{3/2}}{2\mathcal{L}}  + \frac{\Vert F_0 \Vert_{L^2}}{\Vert u(0) \Vert_{L^2}}\right),
\end{align}
and therefore
\begin{align}
    R(N) \sim \mathcal{O}(N^{3/2}),
\end{align}
where we have used the fact that the spectral norm of a discretised function scales as $\mathcal{O}(\sqrt{N})$ and $\Vert \cdot \Vert_{L^2}$ denotes the $L^2$ norm of the continuous function.
Nonetheless, despite the dependency of $R$ on the grid size, from numerical simulation, the linearisation error can be exponentially suppressed even though the sufficient condition, $R<1$ is not fulfilled \cite{Liu_2021}. This further highlights the current inconsistency between discretisation choices and the efficiency of the QCL algorithm.

In principle, we contend that whether an efficient quantum algorithm exists or not for a particular system should depend only on the physics, and nature of the solutions, of the system, not on the discretisation choice. We propose a criterion aimed at establishing a connection between the step size of the discretisation grid, $\Delta x$, and the physical length scales of fluid-flow  dynamics. This criterion poses that surpassing the discretisation scale, which already encapsulates the entirety of the system's physics, is unnecessary. The idea is to find physical parameters that only depend on characteristics of the continuous PDE to determine whether an efficient quantum algorithm exists or not. Crucially, the parameter regime should allow all the physics of the PDE to be resolved. The smallest physical length scale depends only on these physical parameters of the system. The operators $F_2, F_1$, and $F_0$ therefore only depend on these physical parameters. 

\subsection{Length scales}
\label{sec:len_scale}
The dynamics of fluid flows in nature is generally characterized as a highly nonlinear and chaotic phenomenon, in both space and time. The nonlinearity of the flow can be attributed to the emergence of a wide range of length scales that dynamically interact with each other. Such flows are typically governed by the NS equations, given by Eqs.~\eqref{eq:NS1} and~\eqref{eq:NS2}. The specific fluid physics depends on a wide variety of flow parameters and fluid properties, whose collective effect may be studied in a consolidated manner with the aid of non-dimensional numbers.

One such number is the Reynolds number, $Re:=U\mathcal{L}/\nu$, where $U$ and $\mathcal{L}$ are the characteristic velocity and length scales respectively. $Re$ represents the ratio of the inertial to viscous forces in the fluid and its magnitude indicates whether the flow is laminar or turbulent. When $Re<1$, it implies that the viscous forces dominate the inertial ones, leading to enhanced viscous dissipation of the flow energy and thus representing a more laminar-like flow. For $Re\gg1$, the inertial forces dominate the viscous ones, thus paving way for the flow's kinetic energy to translate into potential instabilities that eventually lead to turbulence. Such a transition typically transcends over multiple intermediate mechanisms, before turning into a fully developed turbulent flow, that sets in beyond a threshold $Re=Re_{c}$, whose magnitude typically depends on the specific initial and boundary conditions of the fluid flow.

Therefore, the choice of initial and boundary conditions, together with the instantaneous magnitude of \( Re \), determines the physical scales that characterize the flow state. For instance, the flow could be through a channel or around a bluff body, as shown in Fig.~\ref{fig:turbulence_regime}. In these cases, a characteristic (large) length scale \(\mathcal{L}\) could either be the channel width or the dimension of the bluff body, while \( U \) could represent the channel inlet-flow velocity or the free-stream velocity, accordingly. In fact, even the surface roughness of the boundaries $\delta_{r}$, could also be a relevant length scale, based on the specific physics one intends to investigate.

In turbulence simulations, if the focus is specifically on studying the fundamental ramifications of the nonlinearity in the governing PDE and the various length scales associated, then we can abstract ourselves from the different boundary conditions to some extent, which forms the basis of Direct Numerical Simulations (DNS) of a homogeneous isotropic turbulent flow, in a fully periodic domain. In the \textit{standard DNS set up}, we consider a flow that is constantly energized or stirred at some large length scale $\mathcal{L}$, with a velocity $U$. This corresponds to the forcing or source term $\mathbf{g}$ in the governing PDE. When the forcing is such that $Re$ is sufficiently large, the flow undergoes a chain of instabilities eventually leading to a turbulent flow. 
At this point, a large range of scales is set up in the flow and there exist structures (eddies) that are much smaller in size than the stirring scale itself, see Fig.~\ref{fig:turbine}. Between these large and small scale ranges, there also exists a range called the \textit{inertial range} that might be considered to be more or less independent of effects from the latter scales. As the energy input at large scales gets redistributed among the other existing scales, viscous forces begin to dominate inertial ones around the small-scale structures where local Reynolds number \( Re \sim 1 \), leading to viscous dissipation of the energy being input by the term $\mathbf{g}$. The length scale at which this phenomenon occurs is called the Kolmogorov scale, \(\eta\), and it is typically orders of magnitude smaller than the large scale \(\mathcal{L}\) at high (overall) Reynolds number \( Re \).

When a fluid flow exhibits turbulence, accurately resolving all scales is crucial to uncovering the underlying physics~\cite{iyer2021area}. Consequently, this significantly increases the required discretisation or grid sizes ($N$) of the DNS as the Reynolds number $Re$ grows because the range of scales becomes broader at higher Reynolds numbers. We can make use of the estimates of the smallest length scales that need to be resolved to lower bound the minimum grid spacing required. These bounds can be used to determine whether or not there exists an efficient quantum algorithm that computes the solution with the desired accuracy using the QCL algorithm.

In order to find these estimates, we begin by considering existing theories and state-of-the-art DNS simulations and discuss the bounds on the required grid sizes in three-, two- and one-dimensional flow problems corresponding to the smallest physical scales in each case (all of which, we shall refer to as Kolmogorov length scale), see Fig.~\ref{fig:Summary_dimensions}. We call attention to this (ab)use of terminology in the remainder of the manuscript. 

\subsubsection{Three-dimensional turbulence}
To estimate the Kolmogorov scale in the three-dimensional case, we proceed with the energy redistribution picture by assuming that the rate at which the energy is transferred from the stirring to eddies of different sizes is a constant \cite{pope2000turbulent,davidson2015turbulence}. The characteristic time scale of this process (eddy turnover time) around a large scale eddy could be dimensionally estimated as $\tau \sim \mathcal{L}/U$, with a corresponding energy proportional to $U^{2}$. Therefore, the rate of energy transfer between such scales may be written as, \begin{equation}
    \varepsilon \sim U^{3}/\mathcal{L}.
\end{equation} The eventual dissipation of this energy is a consequence of the dominant viscous effects, that cause shearing and straining of fluid elements at Kolmogorov length and velocity scales, $\eta$ and $u_{\eta}$ respectively. Starting from the NS equations (\ref{eq:NS1},\ref{eq:NS2}), one can compute the strain rate tensors on a fluid element, and thus the rate of energy dissipation as \begin{equation}
    \varepsilon = 2\nu S_{ij}S_{ij},
\end{equation} where $S_{ij}= \frac{1}{2}(\partial u_{i}/\partial x_{j} + \partial u_{j}/\partial x_{i} )$. From dimensional arguments, this quantity may be rewritten as $S_{ij}\sim u_{\eta}/\eta$ and therefore, we get
\begin{equation}
    \varepsilon \sim \nu u_{\eta}^{2}/\eta^{2}.
    \end{equation}
Since $Re = u_{\eta}\eta/\nu \sim 1$, and recalling that rate of energy transfer is constant at all scales, we get $\varepsilon \sim  U^{3}/\mathcal{L} \sim \nu u_{\eta}^{2}/\eta^{2}$. From this, we can readily estimate the Kolmogorov scale to be
\begin{equation}
    \eta \sim Re^{-3/4}\mathcal{L} \sim (\nu^{3}/\varepsilon)^{1/4},
\end{equation}
thus connecting $\eta$ with our non-dimensional quantity, $Re$. We now briefly outline certain essential considerations in resolving these scales in practice using DNS.\\

\noindent \textit{Direct Numerical Simulations} -- This refers to an algorithm that solves the PDEs, in our case the NS equations, directly without any intermediate modeling or dimension reductions. The initial numerical setup could use any of the several existing methods such as finite differences, finite volumes and so on. However, state-of-the-art solvers typically resort to pseudo-spectral methods \cite{rogallo1981numerical}, which we  briefly outline. As a reference, let us consider again, a homogeneous, isotropic turbulent flow in a periodic box of side $\mathcal{L}$, sufficiently large to discard effects from boundary conditions. For the spatial resolution, the box (with $\mathcal{L}$ typically set to $2\pi$), is discretized in every direction with an $N$-point grid, producing a total of $N^{3}$ grid points. 
We now solve Eqs. (\ref{eq:NS1}) and (\ref{eq:NS2}) by first transforming into spectral space using the discrete Fourier transform, 
\begin{align}
    \bm{u}(\bm{x},t) = \sum_{\bm{\kappa}} e^{i \bm{\kappa} \cdot \bm{x}} \hat{\bm{u}}(\bm{\kappa},t).
\label{eq:DFT}
\end{align}
This is motivated by the simplicity of computing derivatives in spectral space by multiplying the spectral velocities by their corresponding wavenumbers $\mathbf{\kappa}$,  that lie in the range $[\mathbf{\kappa_{min}},\mathbf{\kappa_{max}}]\in \big[\frac{2\pi}{\mathcal{L}},\frac{N\pi}{\mathcal{L}}\big] = \big[1,N/2\big]$, where the grid spacing is given by,
\begin{equation}
    \Delta \mathbf{x} = \frac{2\pi}{N} = \frac{\pi}{\kappa_\text{max}}.
\end{equation}
Typically, the largest resolved wavenumber is $ \kappa_\text{max}=2\sqrt{N}/3$, considering the periodicity and aliasing errors of the spectral method. This fact might lead to numerical artifacts in numerical simulations. To accurately resolve the Kolmogorov scale, we therefore require, 
\begin{equation}
    \frac{\Delta\mathbf{x}}{\eta} \equiv \mathbf{\kappa}_\text{max}\eta.
 \end{equation}
By the discussion so far, requiring that $\mathbf{\kappa}_\text{max}\eta = 1$ might suffice to perform an accurate simulation. However, recent efforts \cite{buaria2019extreme} have revealed that $\eta$ actually underestimates the true dissipation length scales. In fact, it is shown that one needs to set at least $\mathbf{\kappa}_\text{max}\eta\geq 6$ in order to resolve the \textit{true} small scales that are characterized by intense velocity gradients and extreme events whose magnitudes lie at least $\mathcal{O}(10)$ standard deviations away from the mean. Such a length scale is shown to scale with $Re$ as, 
 
\begin{equation}
    \eta_\text{ext} = \eta Re^{-\alpha/2}, ~~~~ \alpha = \beta - 1/2,
\end{equation}
where $\beta = 0.78 \pm 0.03$ (is a weak function of $Re$ \cite{buaria2019extreme}). 

A natural question at this point is whether there are going to be even smaller scales. In other words, is there an asymptotic lower bound to the above scaling with $Re$? In Ref. \cite{yakhot2005anomalous}, the authors attempt to answer this question and conjecture that a possible lower bound for the scaling would be given by $\eta_\text{min}\sim Re^{-1} \mathcal{L}$. Therefore, for our purposes, setting $\Delta\textbf{x}<\text{min}\{\eta_\text{ext},\eta_\text{min}\}$ would suffice to perform an accurate DNS simulation.
These estimates particularly relate to an incompressible, homogeneous and isotropic flow. However, upon changing the flow properties and boundary conditions, these estimates of the small scales vary correspondingly. We cite below, two such popularly studied cases.

\noindent \textbf{Effect of compressibility} -- The discussion so far assumes that Eq.~
(\ref{eq:NS2}) holds, i.e. that incompressibilility is enforced, which is characterised by a constant density flow where pressure waves travel through the fluid at the speed of sound ($c_{s}$). However, if we are interested in a compressible flow (represents nearly all flows in nature, where $\nabla\cdot\mathbf{u}\neq0$), the velocity field can exhibit a different, yet small scaled feature in the flow called shocklets, where some flow observables exhibit near discontinuity. In such flows, besides $Re$, the dynamics also depends on another non-dimensional number, known as the turbulent Mach number,
$M_{t} = \sum_{i}\langle u_{i}u_{i}\rangle^{1/2}/c_{s}$. To estimate the smallest length scales, we first note that a compressible flow field can be decomposed into solenoidal ($\mathbf{u}_{s}$) and dilatational ($\mathbf{u}_{d}$) components, such that $\mathbf{u}=\mathbf{u}_{s}+\mathbf{u}_{d}$, where $\nabla\cdot\mathbf{u}_{s}=0$ and $\nabla \times \mathbf{u}_{d}=0$. The solenoidal term dictates the incompressible part of the flow dynamics,  while the dilational component is responsible for the compressible effects. The $\eta$ stemming from the solenoidal part can be estimated similarly as before. The shocklets, on the other hand, typically have a small thickness $\delta$, which is comparable with the corresponding $\eta$. The stronger the shock, the smaller $\delta$. In order to pick the smallest of the two features, we take a look at the distribution function of the ratio $\delta/\eta$. Based on such a distribution, in Refs. \cite{samtaney2001direct,donzis2013relation,jagannathan2016reynolds}, it has been shown that the \textit{most probable} shock thickness, $\delta_{m}$, tends to be of a similar scale as that of $\eta$ and at times can even be a few times larger than $\eta$. This therefore implies, choosing $\Delta x <\eta$ might as well be sufficient to capture all the small scale features of such a compressible flow. \\ 

\noindent \textbf{Turbulent passive scalar mixing} -- We observed that, by just relaxing the incompressibility constraint on the flow, new physical scales (shocklets) of interest emerge, resolving which is important for an accurate simulation. Another such practically relevant scenario could be one that involves a \textit{passive scalar} 
such as a drop of dye added into the flow or scalar temperature field. Here, the question generally is -- how well can turbulence mix a scalar field? \cite{sreenivasan2019turbulent}. This mixing is driven purely by the flow field and the concentration gradient of the scalar. The scalar field undergoes both advection and molecular diffusion, the ratio of which is governed by the Schmidt number, given by $Sc = \nu/D$, where, $D$, is the molecular diffusivity coefficient.
This is termed as the Batchelor scale \cite{batchelor1959small,donzis2008dissipation,donzis2010resolution}, given by, \begin{equation}
    \eta_{B} = \eta Sc^{-1/2}.
\end{equation} Therefore, we can estimate the number of grid points required in this case as,
\begin{equation}
    N^{\text{mixing}}_{3D} \gtrsim Re^{3/4} Sc^{-1/2}.
\end{equation} Although recent efforts \cite{buaria2021small} suggest a slightly gentler scaling, such as $\eta_{D}\sim\eta Sc^{-1/2+\alpha}$,  for our current purposes, the former scaling might suffice.

\subsubsection{Two-dimensional turbulence}
The simulations of such two-dimensional and quasi two-dimensional flows provide significant insights into the study of atmospheric flows and weather prediction. However, stepping the dimensionality of the problem down from three to twodimensions, modifies the physics of the flow quite significantly. This can be seen by considering the  equations for the velocity and vorticity ($\bm{\omega} = \nabla \times \mathbf{u}$), which in two dimensions read as, 
\begin{align}
    \frac{D \mathbf{u}}{D t} &= -\nabla\Big(\frac{p}{\rho}\Big) - \nu\nabla \times \bm{\omega} + \bm{g}(\bm{x}, t), \\
    \frac{D \bm{\omega}}{D t} &= \nu\nabla^{2}\bm{\omega} + \nabla \times \bm{g}(\bm{x}, t).\label{eq:31}
\end{align}
The relevance of $\bm{\omega}$ becomes natural in this discussion   when we note that the energy dissipation can also be expressed as $\varepsilon \sim \nu \langle\omega^{2}\rangle$, where $\langle \cdot \rangle$ represents the ensemble averaging operator. By comparing these equations with their three-dimensional counterpart, Eq. \eqref{eq:31} now lacks the term $(\bm{\omega} \cdot \nabla)\mathbf{u}$, known as the \textit{vortex-stretching} term. In three-dimensional turbulence, this is responsible for intensifying the stretching of vortex structures at small scales and thus the vorticity, eventually leading to dissipation. The absence of this mechanism in two dimensions leads to counter-intuitive phenomena such as inverse-energy cascade, from small to large scales \cite{davidson2015turbulence}. As a consequence of such deviations in the underlying physical mechanisms and flow structures, a different scaling relations emerges, which is now given by, \begin{equation}
    \eta \sim Re^{-1/2}\mathcal{L}.
\end{equation}
Therefore the corresponding grid sizes would scale as
\begin{equation}
    N_{2D} \sim Re^{1/2}.
\end{equation}

\subsubsection{One-dimensional Burgers flow}
In order to study the one-dimension case, we consider Burgers equation,
\begin{align}
    \frac{\partial u}{\partial t} = -  u \frac{\partial u}{\partial x} + \nu \frac{\partial^2 u}{\partial x^2} + f(x,t). 
    \label{eq:Burgers equation}
\end{align}

\begin{figure}[t!]
    \centering
    \includegraphics[width=1\columnwidth]{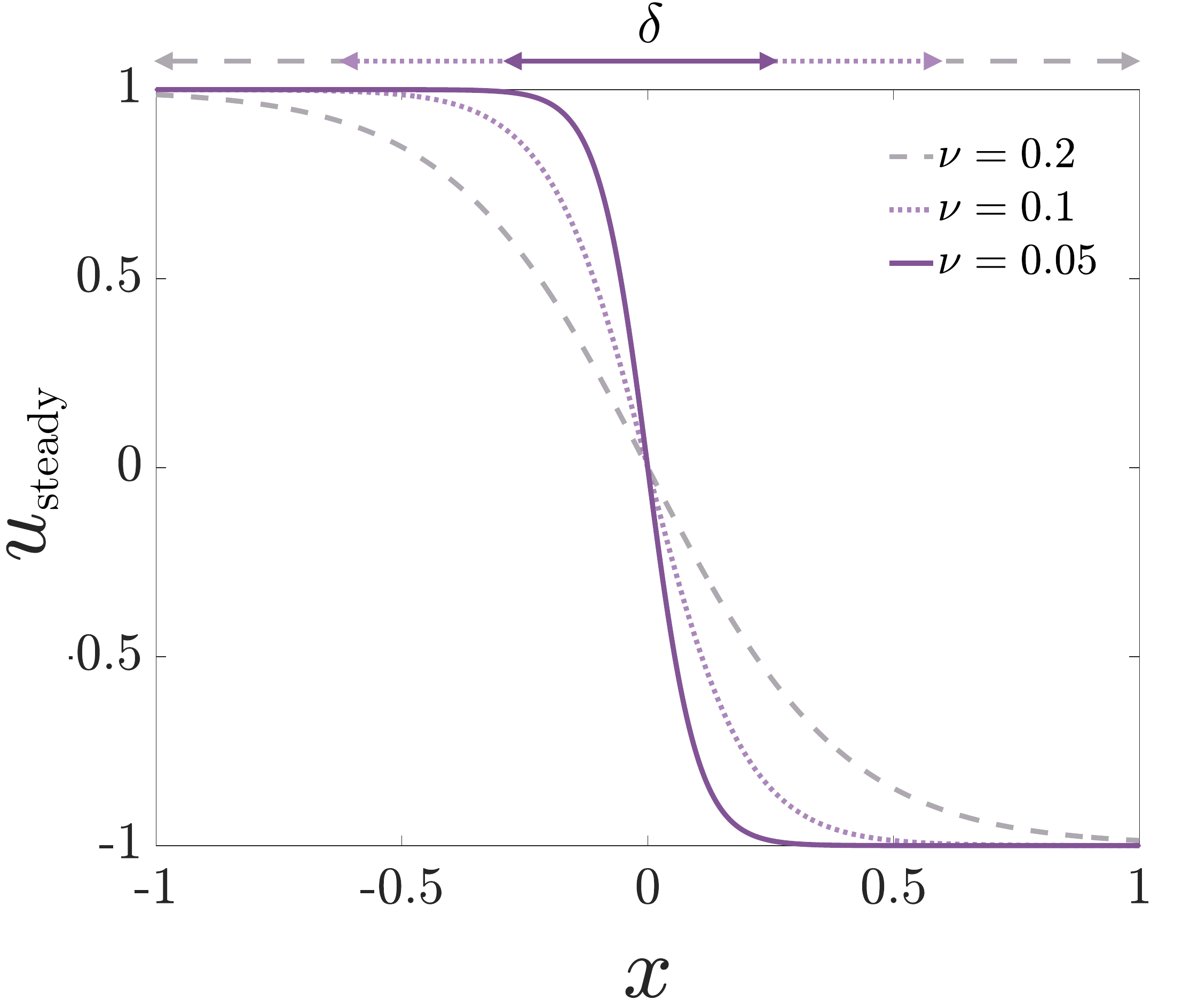}
    \caption{Shock width in the solutions of fluid flows due to highly nonlinear terms in the one-dimensional Burgers equation, Eq. (\ref{eq:Burgers equation}), for different viscosity values. The non linear term term $  -u \frac{\partial u}{\partial x}$  with the Dirichlet boundary conditions $u(-1,t)  = 1$ and $u(1,t)  = -1$ and initial condition $u(x,0) = -x$ has a shocking up effect which becomes stronger as the linear term, $\nu \frac{\partial^2 u}{\partial x^2}$ decreases. This terms generates \textit{shocks} in the solutions whose typical width $\delta$ is given by Eq. (\ref{eq:shock_width}).}
    \label{fig:shock width}
\end{figure}

\begin{figure*}
    \centering
    \includegraphics[width=2.0\columnwidth]{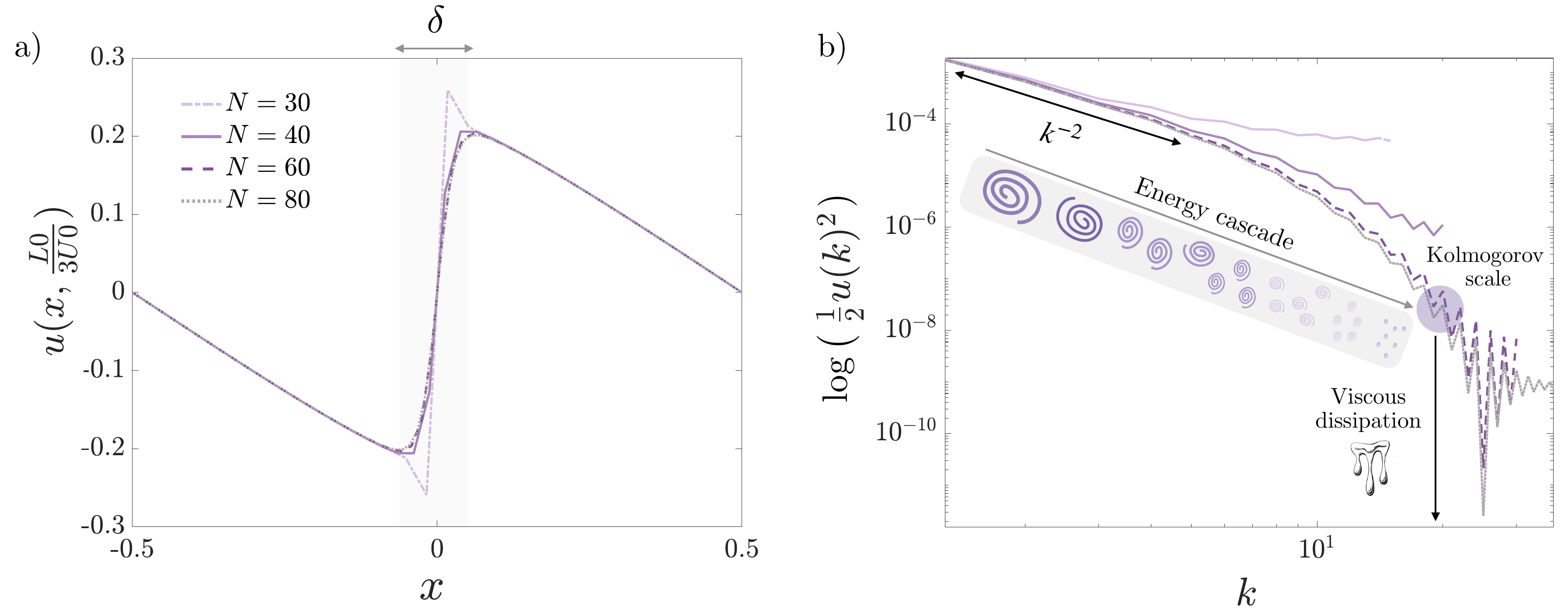}
    \caption{Solutions (a) and  energy cascade (b) for Burgers equation, Eq (\ref{eq:Burgers equation}), in the domain $[\frac{-1}{2},\frac{1}{2}]$ with $Re=80$, Dirichlet boundary conditions $u(-1/2, t)=u(1/2,t)=0$ and initial condition $u(x,0)=\sin (2\pi x)$ for different grid sizes. The energy cascade changes at the Kolmogorov length scale due to the fundamental nature of how energy is transferred from large scales to small scales in turbulent flows. If the grid spacing is too coarse, i.e. larger than the Kolmogorov scale as for $N=30, 40$, the dissipation of energy at these scales will not be properly represented, leading to incorrect predictions of flow dynamics. Furthermore, as the grid spacing decreases to less than the Kolmogorov length scale, there is no significant improvements. Additionally, for large frequencies numerical artifacts might appear in the numerical simulations as already described in \cite{rogallo1981numerical}.}
    \label{fig:numerics_cascade}
\end{figure*}

    \noindent To estimate the ``Kolmogorov scale" in this case, let us consider Eq. (\ref{eq:Burgers equation}) under Dirichlet boundary conditions $u(-1,t) =~u_{a} = 1$ and $u(1,t) = u_{b} = -1$ and an initial condition specified as, $u(x,0) = -x$.

The steady-state solution is as shown in Fig.~\ref{fig:shock width}. The step-like, steep-gradient flow profile, also known as a \textit{shock}, is a signature of the one-dimensional Burgers flow. In contrast to the inviscid case $(\nu=0)$, where the velocity profile exhibits an even sharper discontinuity, a finite viscosity has an effect of spatially \textit{spreading} the shock, giving it a finite thickness. Consequently, at the $Re\gg 1$ regime, there exist several small features in the flow that require accurate numerical resolution \cite{bec2007burgers}. For the present example case, which admits an analytical solution, we may estimate the shock thickness to be $\mathcal{O}(\nu/(u_{b}-u_{a}))$. To see this better, we first observe the shock at $x=0$, which is characterised by the velocity field changing rapidly from $u_{a}$ to $u_{b}$, within the size of the unknown shock width $\delta$. This width can be expressed as
\begin{equation}
\label{eq:shock_width}
    \delta = \frac{u_{b}-u_{a}}{\partial_x u_{|_{x=0}}}.
\end{equation}
Using the analytical solution (available in this special case) given by,
\begin{equation}
    u(x) = \frac{u_{a}+u_{b}e^{\frac{u_{a}-u_{b}}{2\nu}x}}{1+ e^{\frac{u_{a}-u_{b}}{2\nu}x}},
\end{equation}
and the imposed boundary conditions, we can readily estimate the shock width to be,
\begin{equation}
    \delta = \frac{2\nu}{u_{a}-u_{b}}.
\end{equation}
This implies that $\delta$ becomes smaller with decreasing $\nu$ (or increasing $Re$), thus making the shock correspondingly sharper and steeper. Therefore, this phenomenon necessitates an appropriately large grid size such that $\Delta x \leq \delta$, in other words,
\begin{equation}
    N_{1D} \geq \frac{\mathcal{L}}{\delta}\sim Re .
\end{equation} For the rest of the manuscript, we use as reference a slightly modified and extended version of the example discussed above, which we explain shortly in the next section. Nevertheless,  the shock width can be calculated similarly.

\section{Numerical results}\label{sec:num_results}

We now provide some numerical results to illustrate our theoretical analysis. By way of example, we study Burger's equation, given in Eq. (\ref{eq:Burgers equation}) within the domain $[\frac{-1}{2},\frac{1}{2}]$ with Dirichlet boundary conditions $u(-1/2, t)=u(1/2,t)=0$ and  initial condition $u(x,0)=\sin (2\pi x).$

\subsection{Kolmogorov length scale}

We now provide some numerical results on the discretisation size and traceability of physical quantities. In order to resolve the relevant physics, the discretisation must capture all the features of the ``continuous'' solution (so high-powered DNS, imagining that an infinite discretisation is possible). Thus, if the discretisation step is larger than the Kolomogorov length scale, there are features that are not accurately captured. 

In particular, we track the energy cascade of the solution for a Burgers equation with $Re=80$, see Fig.~\ref{fig:numerics_cascade}. The error in the high-powered DNS solver increases as \( \Delta x \) decreases below the value where the grid point spacing is larger than the Kolmogorov length scale. In this sense, if the grid spacing is too coarse, i.e. larger than the Kolmogorov scale $(\Delta x\sim~0.015)$, the dissipation of energy at these scales will not be properly represented, leading to incorrect predictions of te fluid dynamics, as we see for $N=30,\ 40$. Furthermore, as the grid spacing decreases to less than the Kolmogorov length scale ($N\sim 80$), there is no significant improvements. This indicates that we should use Kolmogorov scale discretisation. Thus, the physics, given by the Reynolds number, \( Re \),  determines the discretisation needed to resolve the fluid-flow  dynamics, and therefore, the discretisation is not really a ``choice''.

\subsection{Resolution Efficiency}
We now analyze the efficiency of Carleman linearisation to solve the Burgers equation in combination with the Kolmogorov scale. As Burgers is a one-dimensional equation, we can rewrite Eq. (\ref{eq:R_definition}) as

\begin{align}
    \label{eq:R_definition_Re}
    R(N) \approx \frac{Re}{U\pi^2}\left(\frac{\Vert u(0) \Vert_{L^2}}{2}    N^{3/2} + \frac{\mathcal{L}\Vert F_0 \Vert_{L}^2}{\Vert u(0) \Vert_{L^2}}\right)
\end{align}
By imposing $R=1$ we can describe the \textit{``efficiency frontier''} as 

\begin{align}
    \label{eq:R_frontier}
    Re \approx \frac{U\pi^2}{\left(\frac{\Vert u(0) \Vert_{L^2}}{2}    N^{3/2} + \frac{\mathcal{L}\Vert F_0 \Vert_{L}^2}{\Vert u(0) \Vert_{L^2}}\right)}.
\end{align}
Note that the \textit{efficiency frontier} establishes the limit, in terms of the number of grid points $N$, for those discretisations of dissipative PDEs that can for sure be solved efficiently for any given time via QCL. In the particular case of Burgers equation, Eq. (\ref{eq:R_frontier}) determines the maximum number of grid points for which we can ensure that Eq. (\ref{eq:Burgers equation}) with Reynolds number \( Re \), characteristic velocity scale \( U \), and length scale \( \mathcal{L} \), can be efficiently solved according to Ref. \cite{Liu_2021}.

Additionally, for the one-dimensional case, we can describe the relationship between the Reynolds number and the number of grid points needed to resolve the Kolmogorov scale as
\begin{align}
    \label{eq:Kolmogorov_frontier}
     Re \approx  N.
\end{align}
In the \textit{standard DNS setup}, given a Reynolds number, this equation establishes a boundary for the maximal number of grid points needed, which, \textit{a priori}, does not make sense to exceed, as it will not be possible to resolve new physics beyond the Kolmogorov scale. Note that for higher dimensions, the scaling of $Re$ with $N$ may differ (see Section~\ref{sec:len_scale}). Remarkably, despite the dependence of $R$ on discretisation, we reiterate the crucial point that whether Carleman linearisation converges to an accurate solution of the continuous system dynamics should depend only on the physics of the system e.g. the Reynolds number, and not on the discretisation chosen. of course, assuming that the discretisation is sufficient to capture all the physics. Consequently, from Eq. (\ref{eq:R_definition_Re}), we can define the value of $R$ that corresponds to the Kolmogorov scale as 
\begin{align}
    R^\textrm{KS} \sim \left(  \Vert u(0)\Vert_{L^2} \frac{(2 Re U)^{1.5}}{\pi^2} + \frac{\mathcal{L} \Vert F_0 \Vert_{L^2}}{\pi^2\Vert u(0) \Vert_{L^2}}\right).
\end{align}
Note $R^\textrm{KS}$ depends on the  Reynolds number, which determines the Kolmogorov length scale and, therefore, the  discretisation required to resolve the energy cascade produced by the nonlinear term. Intuitively, by fixing the box size and increasing viscosity, one can reduce the Reynolds number and subsequently the turbulence (nonlinearity) of the fluid.  Therefore, we conjecture that Reynolds number should determine whether the system is linearisable with Carleman truncation or not, as the Reynolds number is the physical quantity in the system that determines its nonlinearity and does not depend on the discretisation.

\begin{figure}[t!]
    \centering
    \includegraphics[width=1\columnwidth]{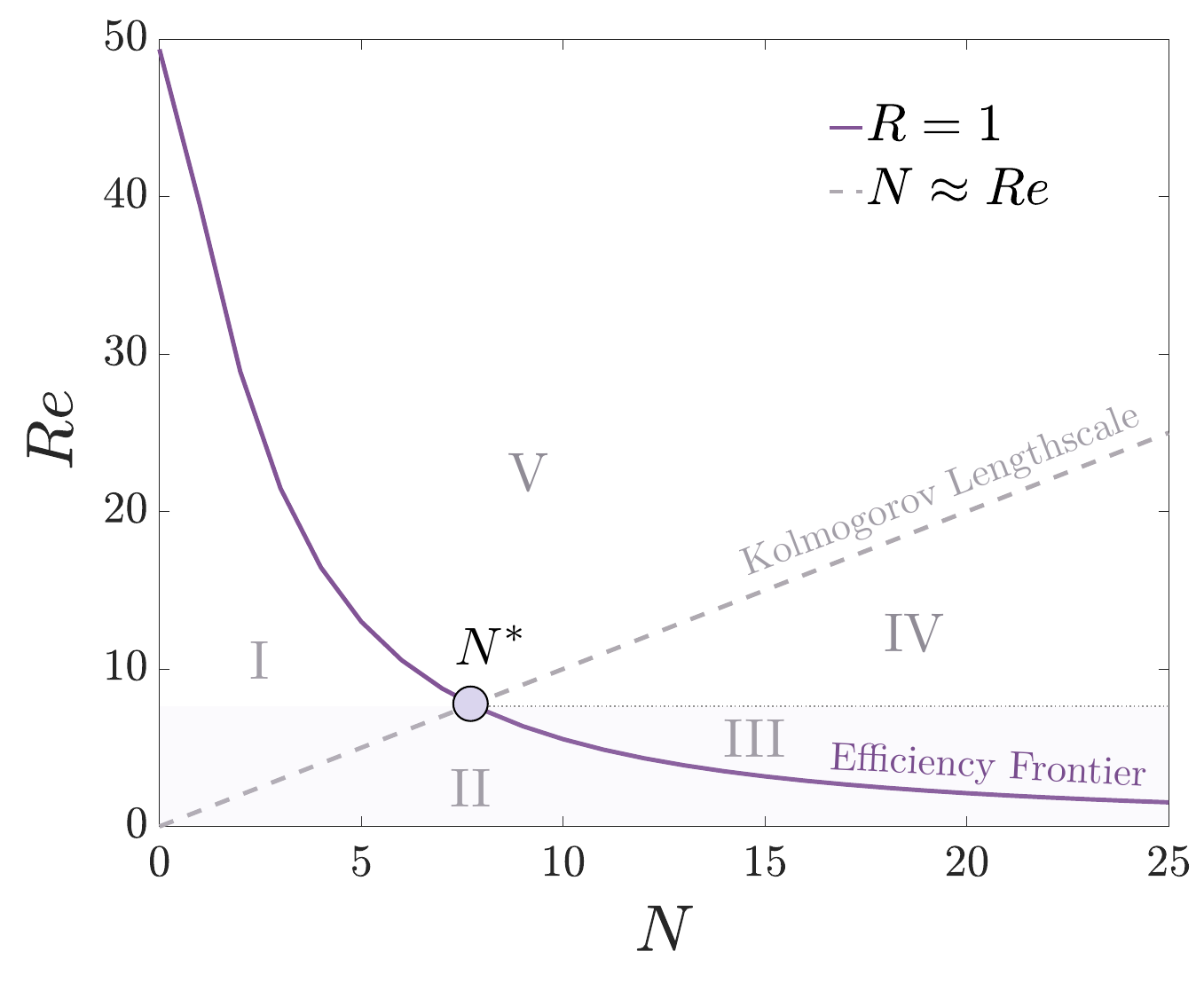}
    \caption{\(N\)-\(Re\) space partition based on efficiency and Kolmogorov scale. The \textit{efficiency frontier} establishes the limit, in terms of the number of grid points $N$, for those discretisations of dissipative PDEs that can be solved efficiently for any given time via QCL, while the Kolmogorov length scale represents the critical minimal scale at which the physics of fluid dynamics can be resolved. Based on these two frontiers we can find the following \(N\)-\(Re\) space partition. 
    Region I: there exists an efficient quantum algorithm but might not resolve all the physics, as the grid is not fine enough.  
    Region II: there exists an efficient quantum algorithm that resolves the physics for scales finer than Kolmogorov. Region  III: the discretisation size is sufficient to resolve all the physics but too large to guarantee efficiency; however, if for the same \(Re\) we choose the Kolmogorov scale, we could resolve Carleman linearisation efficiently.  Region IV: if we coarsen the discretisation to the Kolmogorov scale, efficiency is not guaranteed. Region V: neither efficiency is granted, nor is the discretisation fine enough to resolve all the physics. Wherever $R>1$ we can not ensure there is an efficient algorithm to solve all the physics, but there is numerical evidence for some particular cases \cite{Liu_2021}. Parameters values for the illustrative example: $\mathcal{L}=1$, $U=5$, $\Vert u(0) \Vert_{L^2}=0.5$, $\Vert F_0 \Vert_{L^2}=0.5$. } 
    \label{fig:numerics_1}
\end{figure}

The convergence of Carleman linearisation at the Kolmogorov length scale is guaranteed for all time when $R^\textrm{KS}<1$. For particular systems, convergence with higher $R^\textrm{KS}$ is possible, such as the example in Ref.~\cite{Liu_2021}. We can consider the radius of convergence for Carleman linearisation~\cite{Carleman,amini_carleman_2022, succi2024carlemanroutesquantumsimulation,lin2022koopman}, where up to a time $t^*$ the error converges with $C$ even if $R\geq 1$. We take the example of the differential equation 
\begin{equation}
    \frac{dx}{dt} = - x - \mathcal{R} x^2,
\end{equation}
which is a simplified version of the case in App.~C of Ref.~\cite{lin2022koopman}. In this equation, the parameter $\mathcal{R}$ aligns with $R$ defined in Eq.~\eqref{eq:R_definition}. As shown in Ref.~\cite{lin2022koopman}, the differential equation has a radius of convergence for Carleman linearisation of $t^* = -\log(1 - 1/\mathcal{R})$ for $\mathcal{R}\geq 1$. The error from Carleman linearisation is bounded up to this time. This means, even when $\mathcal{R} \geq 1$, QCL gives an efficient quantum algorithm for simulation times up to $t^*$. For fluid equations, we would expect the time $t^*$ to generally decrease as the Reynolds number becomes larger due to the increasing nonlinearity. If we want to simulate a fluid equation for a time $t^*$, we can define an $R^* > R^\textrm{KS}$ for which Carleman linearisation error is bounded. Simulating times up to $t^*$ would give an \textit{efficiency frontier} that extends into Regions III, IV, and V, increasing the region of applicability of the QCL algorithm.

\begin{figure*}
    \centering
    \includegraphics[width=2.0\columnwidth]{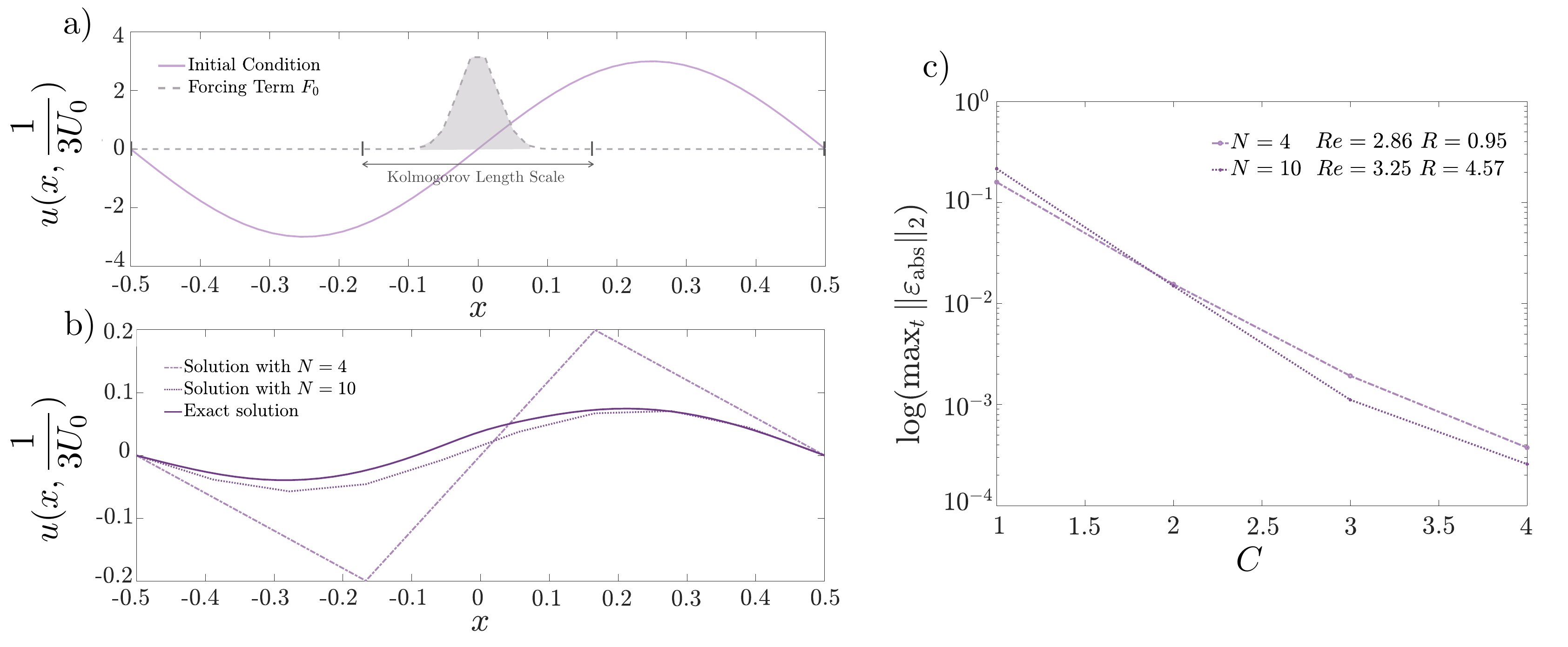}
    \caption{a) Forcing term and initial condition to solve the Burgers equation in a posed set up located in region III. In this example the discretisation corresponding to Kolmogorov length scale does not capture the forcing term because the grid spacing is larger than the length scale of the forcing term.  b) Numerical solutions with infinite number of Carleman terms to the Burgers PDE with $Re=3.3$ for different discretisations at $t=\frac{1}{3U_0}$ with $U_0=\textrm{max}_x |u(x,0)|$, which corresponds to a third of the time the nonlinear effects need topropagate across the whole system. For the $N=4$ case (Kolmogorov length scale) $R=0.95$, while for $N=10$ $R=4.57$. The error to the exact physical (continuous) solution is large, meaning the solution provided at Kolmogorov length scale is not helpful to understand the physics. c) Truncation error with number of Carleman terms $C$ for the solutions with different discretisations sizes.}
    \label{fig:numerics_regionIII}
\end{figure*}

Considering Eqs.~\eqref{eq:R_frontier} and~\eqref{eq:Kolmogorov_frontier}, leads to a partition of the \(N\)-\(Re\) space into five different regions, see Fig.~\ref{fig:numerics_1}. If, for a given \(N\), we can efficiently solve Carleman linearisation for all times but do not resolve the Kolmogorov length scale, we are in region I. If the number of grid points enables an efficient resolution of the problem for length scales finer than the Kolmogorov length scale, we are in region II. We are in region III if the grid size is sufficient to resolve all the physics but too fine to guarantee efficiency; however, if for the same \(Re\) we choose the Kolmogorov scale, we could resolve Carleman linearisation efficiently for all times. If, for the same \(Re\), we coarsen the discretisation to the Kolmogorov scale and efficiency is not guaranteed for all times, we are in region IV. Finally, in region V, neither efficiency is guaranteed, nor is the discretisation sufficiently fine to resolve all the physics.

The most relevant area in this \(N\)-\(Re\) space partition is given by the frontier between regions I and II. Both regions are within the area where Carleman linearisation algorithm is efficient to solve Burgers equation for all times, and the border between the two describes those discretisations that capture the Kolmogorov scale and are efficiently solvable. In this sense, the crossing point between the efficiency and Kolmogorov frontiers is given by the solution to the equation 
\begin{align}
    N = \frac{U\pi^2}{\left(\frac{d}{2}  \Vert u(0) \Vert_{L^2}  N^{3/2} + \frac{\mathcal{L}\Vert F_0 \Vert_{L}^2}{d\Vert u(0) \Vert_{L^2}}\right)}.
\end{align}
We denote this point as \(N^{*}\), and it establishes an a priori estimation of the Burgers equation with the largest $Re$ possible value ($Re\approx N^*$) that can be efficiently solved at the Kolmogorov scale  for all times. Thus, we can conclude that the largest size of the associated linear system of equations that can be efficiently solved is $\mathcal{O}(m(N^{*})^C)$, where $m$ is the number of time steps and $C$ is the truncation order of Carleman linearisation. Therefore, if due to weak nonlinearity (small $Re$), the quantities $C$ and $N^{*}$ are relatively small, such that $(N^{*})^C$ is manageable for a classical computer, a crucial question arises: 
\emph{Is the regime for quantum advantage when solving nonlinear fluid equations with Carleman DNS of practical use?}

In order to address this question, we explore dicretisation sizes beyond $N^*$, which a priori are not granted to be efficiently solved. We conjecture about the efficiency of some problems stated in region III, which are not framed in the \textit{standard DNS setup}. We consider the case of a forcing term on a length scale tinier than the Kolmogorov length scale, i.e. smaller than the dissipation scale due to the viscosity term of the energy from the initial condition. Additionally, we also assume that this forcing is away from the boundaries, to avoid border effects in the fluid-flow  dynamics. As the length scale of the forcing term is shorter, it determines a new physically relevant minimum length scale for the dynamics of the fluid flow. There are two ways in which this could be interpreted: either the forcing term introduces a local (in time as well as space) Reynolds number through the characteristic length scale and velocity of the forcing that gives a smaller Kolmogorov length scale; or the Kolmogorov length scale remains unchanged but no longer captures all the physics of the system due to the smaller length scale of the forcing term -- the physically relevant minimum length scale.

In Fig.~\ref{fig:numerics_regionIII}, we show an example of a choice of  discretisation that does not capture the forcing term, because the grid spacing is larger than the length scale of the forcing term. In this example, the forcing term does not change the Kolmogorov length scale, i.e., a priori, it does not alter the requirements to efficiently capture the nonlinear effects. We find that $R^\textrm{KS}<1$, and therefore for a grid size equal to the Kolmogorov scale the error according to Ref. \cite{Liu_2021} converges exponentially. However, the error to the physical (continuous solution) is large, meaning the solution is not helpful to understand the physics. We observe that, when the discretisation step is decreased such that the grid spacing is smaller than the length scale of the forcing term, the discretised solution from Carleman linearisation becomes physically accurate. Even though $R>1$ in this case, the numerical results show that the error of the QCL algorithm still converges exponentially.

The key point in the example above is that the required discretisation is determined by the system's physics, with a minimum length scale that must be captured, which in this case is defined by the forcing term. Therefore, if the width of the forcing term is smaller than the Kolmogorov scale, one would need to go beyond this scale to be able to accurately capture the impact of this term on the solution. Additionally, choosing a discretisation size larger than the width of the forcing term would result in an homogeneous system of ODEs, leading to solutions with exponential decay that cannot be efficiently solved for long times (see Ref. \cite{Liu_2021}).

The numerical example illustrated in Fig.~\ref{fig:numerics_regionIII}  brings us back to Fig.~\ref{fig:numerics_1}. Given that in this non-standard setup the physically-relevant minimal length scale is smaller than the Kolmogorov length scale, this provides a reason to consider discretisations in regions II and III. In this sense, if the forcing term introduces a smaller minimum length scale but does not alter the Kolmogorov length scale, which is associated with the resolvability of nonlinear effects, one would expect the algorithm to remain efficient even with discretisations finer than the Kolmogorov scale in region III.
Thus, motivated by a source term with a width narrower than Kolmogorov scale, we have observed that  numerically solving equations posed as problems in region III can be efficient, see Fig.~\ref{fig:numerics_regionIII}, even when $R^\textrm{KS}>1$. This motivates us to conjecture that in this set up where the width of the forcing term is smaller than the Kolmogorov scale induced by the initial condition and does not alter it, the problem stated in region III can be efficiently solved.\\

\section{Conclusions}

In this work, we analysed the conditions under which fluid-flow  equations can be efficiently solved using quantum computing. Specifically, we explored the resolution of these equations from a CFD perspective by evaluating the performance of the QCL algorithm in solving DNS. Since the applicability of QCL was originally limited to scalar equations \cite{Liu_2021}, we first extended the formalism to vector fields and presented a detailed numerical framework for this purpose.

Subsequently, in the context of numerical simulations within a \textit{standard DNS setup}, we have shown that the Kolmogorov length scale plays a fundamental role in determining the discretisation size. In this sense, in order to resolve all relevant physics, the discretisation step must be at least as fine as the Kolmogorov length scale. If the grid spacing is larger than this scale, important physical phenomena, particularly the dissipation of energy, can not be accurately captured. Conversely, as the grid size becomes smaller than the Kolmogorov length scale, no significant improvement in accuracy is observed. This indicates that the Kolmogorov length scale represents the critical scale at which the physics of fluid dynamics can be resolved, beyond which further refinement is unnecessary within a DNS standard  setup. 

We also illustrated how the Kolmogorov length scale is strictly connected to the Reynolds number, \(Re\), a key physical quantity that describes the ratio of inertial forces to viscous forces in a fluid flow  and helps to predict fluid-flow  regimes in different physical dimensions. In this sense, the relationship between \(Re\) and the number of grid points, \(N\), that can resolve the Kolmogorov scale is crucial. The higher the Reynolds number, the finer the grid needs to be to capture all the physics. Based on the importance of resolving the fluid flow  at the Kolmogorov scale, we have proposed a connection between the numerical parameter, \(R\), that guarantees efficiency in the truncation of the Carleman linearisation \cite{Liu_2021}, and the Reynolds number, \(Re\). This has enabled us to describe the efficiency of the algorithm in terms of physical parameters with the quantity \(R^\textrm{KS}\), eliminating dependencies on the discretisation size. Moreover, we conjectured that, when \(R^\textrm{KS}\geq 1\), there might exist a convergence radius depending on the weight of nonlinear term that allows for an efficient simulation of QCL up to a certain time $t^*$, expected to decrease as the strength of the nonlinear term increases. Thus, the Reynolds number and the simulation time would determine whether the system is efficiently linearizable with Carleman truncation.

Additionally, the \textit{efficiency frontier} given by $R=1$ and the Kolmogorov length scale have partitioned the \(N\)-\(Re\) space into five distinct regions, each corresponding to different levels of discretisation and computational efficiency. Regions where the discretisation is sufficient to capture the Kolmogorov scale while maintaining computational efficiency at all times are of particular interest, as they represent optimal conditions for solving the PDEs. Conversely, coarser discretisations may fail to resolve the essential physics, leading to inaccuracies, while overly fine grids offer no additional benefit in a standard DNS setup. The intersection between the \textit{efficiency frontier} and the Kolmogorov length scale establishes a maximum number of grid points that should not be exceeded, potentially negating any potential advantage of quantum computers in asymptotic regimes.

Finally, we have studied the role of forcing terms with scales smaller than the Kolmogorov length scale, which introduces a scenario beyond the standard DNS setup. In such cases, the discretisation must capture these finer scales to accurately represent system's dynamics. If the forcing term introduces a smaller length scale, the discretisation must adapt accordingly, even if it exceeds the Kolmogorov scale. Otherwise, the forcing term would effectively be zero, leading to an exponential decay of the solution that impedes the efficiency of the whole quantum algorithm \cite{Liu_2021}. Resolving this scenario presents a challenge but also highlights a situation where efficiency can still be maintained with finer discretisations, even when \(R \geq 1\), and opens the possibility to explore the efficiency of quantum algorithms in asymptotic regimes.

\section*{acknowledgements}
S.S.B would like to thank Katepalli R. Sreenivasan, Dhawal Buaria and John P. John for insightful discussions. The Authors acknowledge support from HORIZON-CL4- 2022-QUANTUM01-SGA project 101113946 OpenSuperQPlus100 of the EU Flagship on Quantum Technologies, the Spanish Ramón y Cajal Grant RYC-2020-030503-I, project Grant No. PID2021-125823NA-I00 funded by MCIN/AEI/10.13039/501100011033 and by “ERDF A way of making Europe” and “ERDF Invest in your Future”, and from the IKUR Strategy under the collaboration agreement between Ikerbasque Foundation and BCAM on behalf of the Department of Education of the Basque Government. This project has also received support from the Spanish Ministry of Economic Affairs and Digital Transformation through the QUANTUM ENIA project call - Quantum Spain, and by the EU through the Recovery, Transformation and Resilience Plan - NextGenerationEU. We acknowledge funding from Basque Government through Grant No. IT1470-22, also from the ELKARTEK program, project “KUBIT—Kuantikaren Berrikuntzarako IkasketafTeknologikoa” (KK-2024/00105).  J.G-C acknowledges the support from the UPV/EHU Ph.D. Grant No. PIF20/276. D.L. acknowledges support from the EPSRC Centre for Doctoral Training in Delivering Quantum Technologies, grant ref.~EP/S021582/1.

\newpage
\bibliography{article.bib}

\end{document}